\documentclass{iopart}
 \expandafter\let\csname equation*\endcsname\relax
 \expandafter\let\csname endequation*\endcsname\relax 
\usepackage{amsmath}
\usepackage{amssymb}
\usepackage{graphicx}
\usepackage{bm}
\usepackage{color}
\newcommand{\be}{\begin{equation}}
\newcommand{\ee}{\end{equation}}
\def\bea{\begin{eqnarray}}
\def\eea{\end{eqnarray}}

\begin{document}
\title{Effective thermodynamics of strongly coupled qubits}
\author{N. S. Williams$^1$, K. Le Hur$^2$ and A. N. Jordan$^1$}
\address{$^1$ Department of Physics and Astronomy, University of Rochester, Rochester, New York 14627, USA} 
\address{$^2$ Department of Physics, Yale University, New Haven, Connecticut 06520, USA}
\ead{nathanw@pas.rochester.edu}
\begin{abstract}
Interactions between a quantum system and its environment at low temperatures can lead to apparent violations of thermal laws for the system.  The source of these violations is the coupling between system and environment, which prevents the system from entering into a thermal state, a state of Gibb's form.  On the other hand, for two-state systems, we show that one can define an effective temperature, placing the system into a `pseudo-thermal' state where effective thermal laws can be applied.  We then numerically explore these assertions for an n-state environment inspired by the spin-boson environment.
\end{abstract}
\pacs{03.65.-w, 03.65.Ta, 05.70.-a}
\maketitle

{\it Introduction ---}
Thermodynamics of small quantum objects, such as two-level systems or harmonic
oscillators, coupled to large environments in the low temperature limit has recently
attracted much attention \cite{MS2,Popescu,Buttner, Mahler} in response to the unprecedented
experimental manipulation of such systems. These studies of the exchange between work and heat in quantum systems is commonly referred to as quantum thermodynamics \cite{QLSN,MS,TK} and asks what happens to work and heat as temperature nears zero; are they still well defined, and do the laws of thermodynamics hold? Many effects arise due to the coupling energy not being the smallest energy scale, as is the case in classical thermodynamics.  As a matter of fact, the strong interaction (where we use the term strong to indicate that the weak interaction limit no longer applies) between
the system and the bath prevents the system from entering
into a thermal state (a state of Gibb's form), which for $T=0$ implies entanglement but for finite $T$ is a more subtle issue \cite{Nagaev2002,Jordan2,Jordan,Hilt2009,LeHur2009,Hilt2010}. Therefore, no well-defined local temperature exists
associated with the system itself, and naively applying the equilibrium temperature
of the total system may result in apparent violations of the second law of
thermodynamics \cite{AEA,AEA2} in the form of the Clausius Inequality (CI) \cite{Thermo,Mahler}, and violations of Landauer's Principle (LP) \cite{Landauer,Buttner}.  While there has been previous work on this topic \cite{Ford1985,Ford2005,Ford2006,Hanggi2006,Campisi2009} which provides a solution using a Helmholtz free energy calculated from the total free energy minus the free energy of the bath in absence of the system, we specifically address a system commonly used as a model in low energy physics and take a different view point from previous work keeping the Helmholtz free energy calculated for the system alone, but in an effective framework.
We investigate the quantum thermodynamics of a two-level system (or spin-1/2) coupled to an environment, which has many applications from
dissipative quantum mechanics to limitations in quantum computing \cite{Buttner,Bennett1995,Ford2005}, such as heat dissipation during quantum algorithms.

We show that at any temperature the spin reduced density matrix can be 
thought of
as an uncoupled two-level system at an effective temperature 
$T^*$.  Using this effective framework, we check
that the second law of thermodynamics involving $T^*$ will hold, the 
CI becomes in fact a strict equality and LP is satisfied. The idea of an effective local temperature was also mentioned in \cite{Weimer2008} where they redefine the quantum quantities of work and heat in terms of whether energy transfers change the entropy of the system, while our effective framework utilizes definitions that have been carried over from classical thermodynamics.  We also show that as temperature increases, our bath behaves as a purely thermal bath (with $T^* \rightarrow T$ as the thermal energy scale nears the coupling energy), where
the von Neumann entropy essentially converges to the thermal entropy.

{\it Model ---}  
The system under investigation is a two-level qubit coupled to a bath in the quantum limit (close to absolute zero) modeled by a Hamiltonian of the form $\hat H = \hat H_S + \hat H_B + \hat H_{int}$ where it has been divided into system, bath, and interaction components. 
We start by assuming that the total system (composed of the two-level system described by a Hamiltonian of $H_S = \frac{1}{2}( \Delta \hat\sigma_X + \epsilon \hat\sigma_Z)$ and the thermal reservoir) is in a thermal state at temperature $T = 1 / \beta$ (we set $\hbar = k_B=1$ and for simplicity),
\be
\rho_{tot} \propto {\rm exp}(- \beta \hat H).
\ee
All information about the system itself is contained in the reduced density matrix with the bath degrees of freedom traced out, $\rho_s = {\rm Tr}_b [\rho_{tot}]$.  Using this reduced density matrix the energy fluctuations can be calculated at a given temperature ($\langle \delta \hat{H}_S^2 \rangle = \langle \hat{H}_S^2 \rangle - \langle \hat{H}_S \rangle^2$) and found to increase with the coupling between system and environment in agreement with the work done by Nagaev and B\"uttiker \cite{Nagaev2002}.  For a finite temperature, or zero temperature, and a given bath and interaction this reduced density matrix can be difficult to calculate, so we calculate with a general two-level system density matrix,
\be
\rho_s = \frac{1}{2}
\begin{pmatrix}
1 + Z && X - i Y \\
X + i Y && 1 - Z
\end{pmatrix}.
\ee
Any two-level state can be represented in this way with $X$ ($Y$,$Z$) being the expectation values of the Pauli matrices ${\rm Tr}[\hat\sigma_X \rho_s]$ (${\rm Tr}[\hat\sigma_Y \rho_s]$, ${\rm Tr}[\hat\sigma_Z \rho_s]$).  For simplicity, we assume the system is not coupled via the $\hat \sigma_Y$ operator, allowing $Y$ to be set to zero \cite{footnote1}. 

First we diagonalize the system density matrix.  The eigenvalues of the reduced density matrix are
\begin{equation}
\lambda_{\pm} = \frac{1}{2}(1 \mp P ), \hspace{.5cm}  P = \sqrt{X^2 +Z^2},
\end{equation}
with respective normalized eigenvectors of
\be
|\phi_{\pm}\rangle = \sqrt{\frac{1}{2P(P \pm Z)}}\begin{pmatrix}X \\ -Z \mp P \end{pmatrix} .
\ee
Physically we can interpret the $\lambda_{\pm}$ as the probabilities of being in the $|\phi_+\rangle$ or $|\phi_- \rangle$ state, respectively.
Now, we define an equivalent framework in terms of an uncoupled
two-level system at an effective temperature $T^*$, such that $\rho_s \propto
{\rm exp}(-\beta^* H_S^*)$ and $H_S^*$ of the uncoupled two-level
system is defined below.  With this effective temperature we treat the system as if it were a thermodynamic entity obeying effective thermal laws.

We define an effective ``uncoupled'' Hamiltonian in order to determine parameters such as the effective heat.  In order to do this we assume the effective Hamiltonian is that of an uncoupled two level system,
\be
\hat H_S^* = \frac{1}{2} (\Delta^* \hat\sigma_X + \epsilon^* \hat\sigma_Z),  
\ee
where $\Delta^*$ and $\epsilon^*$ which are the effective tunneling energy and level asymmetry, respectively.
The associated energy eigenvectors of $H_S^*$ are
\begin{equation}
\begin{gathered}
| \phi_{\pm}^{*} \rangle = \sqrt{\frac{1}{2 \eta (\eta \mp \epsilon^*)}}\begin{pmatrix}\pm \Delta^* \\ \eta \mp \epsilon^* \end{pmatrix},
\end{gathered}
\end{equation}
with $\eta = \sqrt{\epsilon^{*2}+\Delta^{*2}}$,
which correspond to effective energy eigenvalues of
\begin{equation}
E_{\pm}^* = \pm \frac{1}{2} \eta.
\end{equation}
We assume these eigenvectors are also eigenvectors of a pseudo-thermal state in Boltzmann form allowing a comparison between these states and energies to the coupled case.  From that comparison we find the effective parameters ($\Delta^*$, $\epsilon^*$, and $T^*$) to treat the system as an uncoupled two-level system in pseudo-thermal equilibrium.

We have three unknowns to determine ($\Delta^*$, $\epsilon^*$, and $T^*$).  Identifying the eigenvalues of the reduced density matrix (3) as weights from a Boltzmann distribution for the pseudo-thermal state, the effective temperature can be written in terms of the other two variables,
\begin{equation}
T^* = \sqrt{\epsilon^{*2} + \Delta^{*2}}/ \ln [\lambda_+ / \lambda_- ] ,
\end{equation}
which must be greater than or equal to zero from the definitions of energy (7).
Due to the freedom given to us by introducing the three effective framework parameters ($\Delta^*$, $\epsilon^*$, and $T^*$) we are allowed to make a choice regarding the expectation value of $H_S^*$.
Continuity can be promoted by choosing that the expectation value of the two-level system energy be the same in the normal and effective framework ($\langle \hat H_S \rangle = \langle \hat H_S^* \rangle$), obtaining the relation
\begin{equation}
\epsilon Z + \Delta X = \epsilon^* Z + \Delta^* X.
\end{equation}
It should be noted here that one can calculate the additional moments of the system Hamiltonian following a method similar to Jordan and B\"uttiker \cite{Jordan2} and find that this choice does not guarantee the equality of the other moments (in general $\langle \hat H_S^n \rangle \neq \langle \hat H_S^{*n} \rangle$ where $n\ge2$) highlighting a weakness of this formalism previously undiscussed in the literature.  Specifically for this case $\langle \hat{H}_S^2 \rangle = (\epsilon^2 + \Delta^2)/4$ and $\langle \hat{H}_S^{*2} \rangle = (\epsilon^{*2} + \Delta^{*2})/4$, which are not equivalent unless the system is in a completely mixed state with $X = Z = 0$.

Another relation for the effective parameters arises from the eigenvectors.  We require the eigenvectors of the effective uncoupled two-level system (6) to be equivalent to the eigenvectors of the coupled system (4) corresponding to being in the excited (ground) state of $H_S^*$.  We accomplish this by defining the effective parameters such that the eigenvectors have the same direction in Hilbert space.  In order to determine these definitions we set the ratio of the elements of $ | \phi_{+} \rangle$ ( $ | \phi_{-} \rangle$ ) equal to the ratio of elements from $| \phi_{+}^* \rangle$ ( $| \phi_{-}^* \rangle$ ) and find
\begin{equation}
\begin{gathered}
\eta = -\frac{\Delta^*}{X}(P - Z) - \epsilon^* \hspace{.5cm} \text{From $\phi_-$ and $\phi_-^*$},\text{ and}\\ \\
\eta = -\frac{\Delta^*}{X}(P + Z) + \epsilon^* \hspace{.5cm} \text{From $\phi_+$ and $\phi_+^*$}.
\end{gathered}
\end{equation}
Setting these equations equal we can solve for $\epsilon^*$,
\begin{equation}
\epsilon^* = \Delta^* \frac{Z}{X}.
\end{equation}
We can now use the three Eqs. (8,9,11) to solve for $\epsilon^*$, $\Delta^*$, and $T^*$.
\begin{equation}
\begin{gathered}
\epsilon^* = \frac{\epsilon Z + \Delta X}{X^2 + Z^2} Z, \hspace{.5cm} \Delta^* = \frac{\epsilon Z + \Delta X}{X^2 + Z^2} X, \\
T^* = \frac{| \epsilon Z + \Delta X |}{ P \ln [\frac{1+P}{1-P}]} = -\frac{\epsilon Z + \Delta X}{ P \ln [\frac{1+P}{1-P}]},
\end{gathered}
\end{equation}
requiring $T^*\ge 0$ from the definition of $T^*$ (8) and knowing that $\epsilon Z + \Delta X \le 0$ \cite{footnote3}.  The effective temperature as a function of the physical temperature can be seen in Fig. 2, the effective epsilon and delta do not change much within our temperature scale under the assumption $\epsilon \ll \Delta$ used in the numerical analysis. 
Such an effective temperature has been defined for other quantum objects such as an oscillator \cite{Mahler}.
Now that we have our effective parameters we have the effective Hamiltonian of the system, $\hat H_S^* = \frac{1}{2} (\Delta^* \hat\sigma_X + \epsilon^* \hat\sigma_Z)$, with the system in the ``equilibrium'' described by $\rho_S \propto {\rm exp}(- \beta^* H_S^*)$.
This effectively uncoupled two level system can be treated as if at a temperature $T^*$ and recover the reduced density matrix of the coupled system, as well as the expectation value of the energy.  It should be noted that as the temperature $T$ increases, the interaction between system and bath should play less of a role and the system's reduced density matrix will tend towards one of a thermal state at temperature $T$.  This will result in the effective temperature converging with the physical temperature as the thermal energy scale nears the coupling energy.

We now have an effective temperature and can examine the CI which states $\delta Q \le T d S$, with $\delta Q$ being the heat transferred, $T$ is the temperature, and $d S$ is the change in the entropy of the system.  We examine the system in terms of the effective forms of the heat, temperature, and entropy.  With the effective temperature calculated we can now look at the effective thermodynamic entropy.  We define this entropy as the negative derivative of the Helmholtz free energy with respect to temperature, $S(T^*) = - \partial F / \partial T^*$, and the Helmholtz free energy as $F = - T^* \log Z$, with $Z$ being the partition function of the system alone, $Z = e^{-\beta^* E_-^*} + e^{-\beta^* E_+^*}$, where $E_-^*$ and $E_+^*$ represent the ground and excited energies of the two-level system, $H_S^*$.
Since a thermal state is diagonal this effective thermodynamic entropy is equivalent to the system von Neumann entropy, $S = -  \displaystyle\sum_i \lambda_i \ln \lambda_i$ \cite{NCbook}
\begin{equation}
\begin{gathered}
S = - \frac{\partial}{\partial T^*} \Big[ - T^* \ln [ e^{-\beta^* E_-^*}+e^{-\beta^* E_+^*}] \Big] \\
= -  \displaystyle\sum_i \lambda_i \ln \lambda_i = S_{\nu},
\end{gathered}
\end{equation}
where $\lambda_{\pm} = e^{-\beta^* E_{\pm}^*} / [ e^{-\beta^* E_-^*}+e^{-\beta^* E_+^*}]$.
Inserting the eigenvalues (3), this becomes
\begin{equation}
S = \ln 2 - \frac{1}{2} \ln \Big[(1+P)^{1+P}(1-P)^{1-P} \Big].
\end{equation}
Next we calculate the effective heat of the system.  We go back to the definition of the heat as $\delta Q^* = {\rm Tr}[H^* \delta \rho ]$, which leads to
\begin{equation}
\delta Q^* = \frac{1}{2} \Big( \epsilon^* \delta Z + \Delta^* \delta X \Big).
\end{equation}

Now we know all the terms for the CI we can look at how they change and examine the validity of this effective CI.
Combining the effective temperature with the change in entropy we find
\begin{equation}
T^* d S =  \frac{1}{2P}( \epsilon Z + \Delta X ) \delta P.
\end{equation}
We can now write out the effective CI in terms of the differentials
\begin{equation}
\begin{gathered}
\delta Q^* \le T^* d S, \\
\frac{1}{2}(\epsilon^* \delta Z+\Delta^* \delta X ) \le  \frac{1}{2P} ( \epsilon Z + \Delta X ) \delta P.
\end{gathered}
\end{equation}
Simplifying this equation and substituting back in for $P,\epsilon^*$, and $\Delta^*$ we end up with
\begin{equation}
\begin{gathered}
\frac{\epsilon Z + \Delta X}{X^2 + Z^2} Z \delta Z + \frac{\epsilon Z + \Delta X}{X^2 + Z^2} X \delta X \\ \\ \le  ( \epsilon Z + \Delta X ) \frac{X \delta X + Z \delta Z}{X^2+Z^2},
\end{gathered}
\end{equation}
finding that the effective CI is identically an equality,
\begin{equation}
\delta Q^* = T^* d S.
\end{equation}
When we consider the effective temperature as the correct thermodynamical quantity to be used with thermodynamical laws, then there are no violations.

{\it Numerical Simulations ---}
To further our exploration of the thermodynamics of these systems we conducted numerical simulations.
We modeled a qubit linearly coupled to a bath of harmonic oscillators with $M$ finite modes, each with only the $N$ lowest energy levels.  We define our Hamiltonian according to the spin-boson model \cite{SBM,kondo,lehur} 
\be
\hat H = \frac{\Delta}{2} \hat\sigma_X + \frac{\epsilon}{2} \hat\sigma_Z + \displaystyle\sum\limits_{i} \omega_i \hat a_i^{\dagger} \hat a_i + \frac{1}{2} \hat\sigma_Z \displaystyle\sum\limits_{i} \lambda_i (\hat a_i^{\dagger} + \hat a_i).
\ee
and impose a finite number of modes, truncated after the $N$ lowest levels.  The harmonic oscillator bath is described by $\omega_i$, the frequency for the i$^{th}$ mode, and the creation and annihilation operators $\hat a_i^{\dagger}$ and $\hat a_i$.  The interaction is linear and controlled by the coupling parameters, $\lambda_i$.  We are dealing with low temperatures so most of the population of the bath is contained in the lower levels giving merit to this approximation.  We set the combined bath + system to a thermal state at temperature $T$ and trace out the bath environment to get the reduced density matrix of the system.  This allows the computation of the system expectation values.

For our examination we set our parameters in the regime of $\epsilon \ll \Delta$ (specifically $\epsilon = 0.01 \Delta$ for the results presented), and varied the temperature between $T=0$ and $T=  \Delta$ as well as the coupling constant, $\alpha$.  We define $\alpha$ such that $\lambda_i = \sqrt{2 \alpha} \omega_i$, this definition of $\alpha$ is inspired by the spectral density function of the Spin Boson model\cite{SBM} but modified to account for a bath of finite discrete modes in such a way that when we take the number of modes to infinity we recover the Spin Boson density.  We investigate the simple case of a single mode bath and truncate to only two energy levels in the mode (more complex baths exhibit similar behavior as this basic case).
All of the information needed to examine the system is contained in the $X$, $Y$, and $Z$ expectation values (for our system $Y=0$ \cite{footnote1}).  
We find that as the temperature increases the system becomes more mixed and the expectation values tend towards zero, representative of becoming a fully mixed state as expected for thermal energies approximately equal to the system energy scales, as expected from previous work \cite{lehur}.  

With these expectation values we can examine the entropy of the system, both the von Neumann entropy as well as the thermodynamic entropy based on the Helmholtz free energy as before, but using the actual temperature and energy as opposed to the effective ones.  The von Neumann entropy is defined as in Eq (14).
Plots of these two entropies can be seen in Fig. 1.

%Plots of Entropies (both thermal and von Neumann)
%%%%%%%%%%%%%%%%%%%%%%%%%%%%%%%%%%%%
\begin{figure}[t]
\begin{center}
\includegraphics{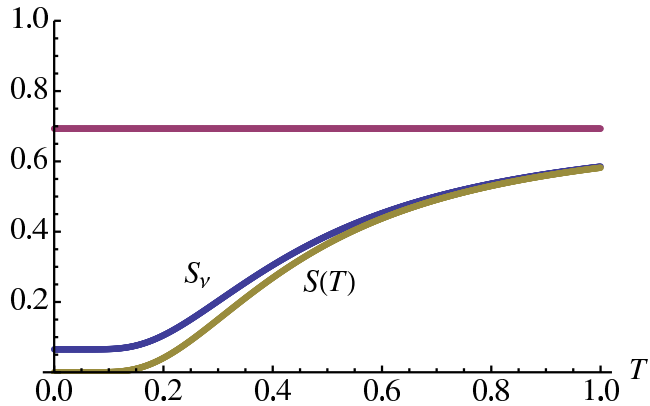}
\caption{(Color Online) Thermal entropy, $S(T)$ and von Neumann entropy, $S_{\nu}$ as a function of temperature, $T$, in units of $\Delta$.  The purple horizontal line signifies the maximum entropy of a two-level system ($S = \log 2$).  For this data $\alpha$ = 0.1.}
\label{fig 2}
\end{center}
\end{figure}
%%%%%%%%%%%%%%%%%%%%%%%%%%%%%%%%%%%%

Mapping the spin boson model to the anisotropic Kondo model \cite{lehur}, we find that the entropies coincide when $T \approx T_K$ (the Kondo temperature) and the reservoir acts as a thermal bath.  This is a similar result as found by H\"orhammer and B\"uttner \cite{Buttner} for a harmonic oscillator chain.  The deviation between the entropies near $T=0$ is a result of the interaction between system and bath, which entangles the system and bath causing the system to be in a mixed state, and therefore $S_\nu \ne 0$.  While $S(T) \rightarrow 0$ at $T=0$, which can be understood by analogy to the anti-ferromagnetic anisotropic Kondo model \cite{lehur,SBM,kondo} where the spin is ``fully screened'' at low temperatures. This leads to violations of LP as can be seen in Fig. 2.

{\it Landauer Principle ---} As in the work of H\"orhammer and B\"uttner \cite{Buttner} we can look at how this quantum two-level system relates to the LP \cite{Landauer}.  We can define our differential of heat as $\delta Q = T d S(T)$ and look at the ratio $\delta Q / d S_{\nu}$, where we vary these parameters with respect to temperature.  The LP states that the ratio of heat released to change in entropy should be greater than or equal to $ T$,
\begin{equation}
\delta Q / d S_{\nu} \ge T.
\end{equation}
Physically the increase in entropy is interpreted as erasure of information.  In our system, full erasure corresponds to a change in entropy of $\log 2$ (from Eq. (14)).
We can plot this ratio for our two-level system coupled to a thermal bath and compare it to this inequality (Fig. 2).
%%%%%%%%%%%%%%%%%%%%%%%%%%%%%%%%%%%%
\begin{figure}[t]
\begin{center}
\includegraphics{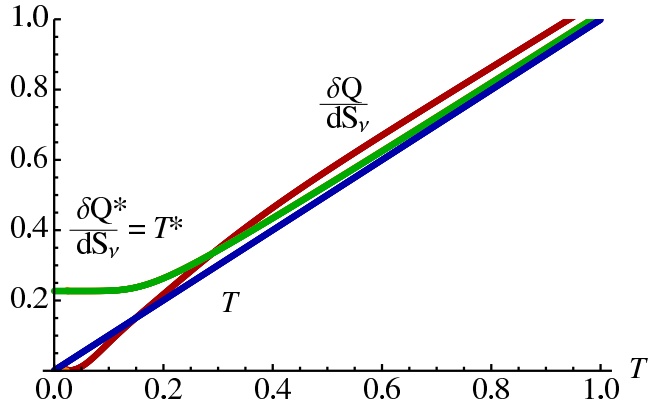}
\caption{(Color Online) The differences between the LP seen normally and in the effective framework.  One can see that there is an apparent violation when looking at the change in heat over change in entropy, $\delta Q/ d S$, but switching to the effective heat results in no violation.  Note that $T^*$ converges to $T$ as $T$ increases.}
\label{fig 3}
\end{center}
\end{figure}
%%%%%%%%%%%%%%%%%%%%%%%%%%%%%%%%%%%%
\begin{figure}[ht]
\begin{center}
\includegraphics{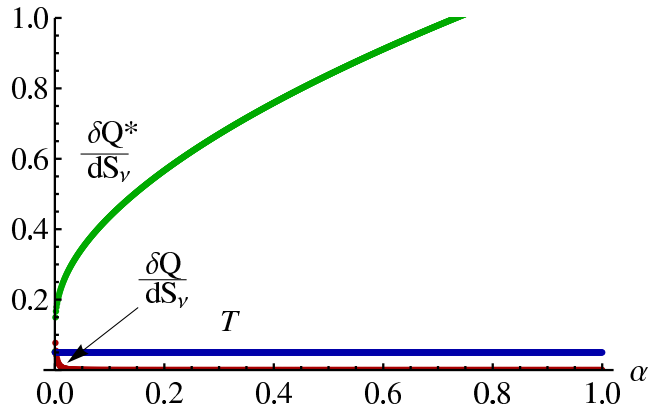}
\caption{(Color Online) The differences between the LP seen normally and in the effective framework.  This shows the violation as a function of the parameter $\alpha$.  The temperature is set at $T = 0.05 \Delta$ for this plot.}
\label{fig 4}
\end{center}
\end{figure}
%%%%%%%%%%%%%%%%%%%%%%%%%%%%%%%%%%%%%
From this plot we can see that at low temperatures there is a violation of the LP, very similar in behavior to what was seen for the harmonic oscillator system \cite{Buttner} showing that this behavior may be a general result for interacting quantum systems.

Following the procedure put forth by Kim and Mahler \cite{Mahler} and discussed above we can define an effective heat and look at the LP in terms of these new quantities (Fig. 2).
The LP ends up being an equality in this effective framework, as suggested by the analytical solution.  This approach can restore the validity of the LP, but only at the cost of introducing an elevated effective temperature, originating from the strong environmental interaction.

{\it Conclusion ---} We have shown that for a two-level system coupled to any bath one can define an effective temperature and place the system into a pseudo-thermal state.  At low temperatures, the strong interaction between the system and environment prevents the system from entering into a thermal state leading to violations of thermal laws for the system, though the overall state is thermal (even at $T=0$).  Even though there are some weaknesses to this effective framework (higher moments of the Hamiltonian not being equivalent), once we define this pseudo-thermal state there are no violations of the CI or LP and the system should follow according to the effective laws of thermodynamics.  At high temperatures \cite{lehur}, this effective framework converges with the classical thermodynamics.

KLH acknowledges NSF through the grant DMR-0803200 and through the Yale Center for Quantum Information Physics (NSF DMR-0653377).  NSW and ANJ acknowledge the support from the NSF through the grant DMR-0844899.


\begin{thebibliography}{99}
\bibitem{footnote1}
We can make a simplification from a symmetry argument.  Since the Hamiltonian is symmetric under the transformation $\hat\sigma_Y \rightarrow -\hat\sigma_Y$ we can set $Y = 0$ for our system.
\bibitem{footnote3}
The fact that $\langle \hat{H}_S \rangle = \epsilon Z + \Delta X \le 0$ arises from the fact that a thermal state will not have any inversion for positive temperatures.  Because a thermal state (one of Gibbs form) has weights of the form ${\rm exp}[- \beta E]$ the ratio of populations between the excited and ground state of the two level system will be ${\rm exp}[- \beta (E_e-E_g)]$ which will always be less than or equal to 1.  Because of this fact when the zero point for energy is set at $(E_e+E_g)/2$ the expectation value of the energy of the system will always be less than or equal to zero for $T \ge 0$.
\end{thebibliography}
\end{document}